# Fermi surface magnetization of Fe-doped NbSb$_2$ investigated by magnetic quantum oscillations

Sang-Eon Lee[1], Sanghyun Ji[1], Myung-Hwa Jung[1,*]

[1]*Department of Physics, Sogang University, Seoul 04107, Korea*

Magnetic quantum oscillations (MQOs) have been widely used as a tool for probing Fermi surfaces. The shape and topology of the Fermi surface and the related physical parameters, such as the cyclotron mass and relaxation time, can be verified by carefully analyzing the frequency, amplitude, and phase of MQOs. In particular, phase analysis, from which we can obtain the Berry phase, has received much attention with the growing interest in the topology of condensed matter physics. Here, beyond the conventional uses of MQOs, we show that MQOs can be used to determine the Fermi surface magnetization. We doped dilute magnetic element Fe into the Dirac semimetal NbSb$_2$ to only introduce magnetism without changing the shape of the Fermi surface. We observed a phase shift in magnetically doped Fe-NbSb$_2$, which is in contrast to the lack of a phase shift in pristine NbSb$_2$ and nonmagnetic Bi-doped NbSb$_2$, indicating the strong exchange interaction between doped magnetic impurities and Fermi surface electrons. We estimated the Fermi surface magnetization introduced by magnetic Fe doping from the phase shift. This work demonstrates not only how tiny magnetic impurities can significantly change the Fermi surface magnetization but also how the Fermi surface magnetization can be investigated by phase analysis of MQOs.

# I. INTRODUCTION

Magnetic quantum oscillations (MQOs) are a useful tool used for investigating the properties of the Fermi surface by measuring the resistance (Shubnikov-de Haas effect) [1–3] or magnetization (de Haas-van Alphen effect) [4,5] with varying applied magnetic field. The frequency and phase of MQOs are directly related to the quantization condition of the Fermi surface electron, which is determined by the cross-section area and topology of the Fermi surface [6–8]. The amplitude of MQOs is well explained by the Lifshitz-Kosevich (LK) formula, which treats the effect of finite temperature and relaxation time as Fermi level broadening [9]. From the well-established LK formula, the fermiology of various materials has been verified by analyzing MQOs. Recently, phase analysis of MQOs has drawn much attention with the growing interest in topological materials. The nontrivial topology of topological insulators [10–14] and Dirac semimetals [15–21] has been investigated by phase analysis of MQOs.

The orbit topology of topological materials has been studied by analyzing the phase, $\lambda = \phi_B + \phi_R + \phi_Z$, where $\phi_B$ is the Berry phase, $\phi_R$ is the Roth phase from the orbital magnetization, and $\phi_Z$ is the phase from the spin magnetization related to the Zeeman coupling. For massless Dirac or Weyl semimetals (negligible $\phi_R$) with small Zeeman coupling ($m_c << m_e$), the Fermi surface topology has been well understood by the dominant contribution of $\phi_B$ [10,19,22–25]. Recently, this method has been generalized to massive Dirac semimetals with considerable cyclotron mass ($m_e \sim m_c$), incorporating $\phi_R$ and $\phi_Z$ [19]. In our previous report on the three-dimensional massive Dirac semimetal NbSb$_2$, we verified the orbit topology by observing an abrupt π phase shift due to the interference between the spin-degenerate orbits.

As mentioned above, $\lambda$ is related to the magnetic property of the Fermi surface. However, the attempt to measure the Fermi surface magnetization by measuring MQOs has never been made. Here, we established the method for measuring Fermi surface magnetization via $\lambda$, which can be applied to general magnetic materials from paramagnets to ferromagnets having their Fermi surfaces. And we applied the method to a dilute magnetic semimetal, Fe-doped NbSb$_2$. We measured the dHvA effect of Fe-doped NbSb$_2$, and analyzed the phase of MQOs. To avoid affecting the shape of the Fermi surface, we doped a tiny amount of magnetic Fe and compared all the results with those of pristine NbSb$_2$ and nonmagnetic Bi-doped NbSb$_2$. We observed an unusual phase shift for Fe-doped NbSb$_2$ and no phase shift for NbSb$_2$ and Bi-doped NbSb$_2$. To explain the phenomenon, we introduced the Fermi surface magnetization that comes from the interaction between conduction electrons and localized magnetic moments and quantitatively estimated the Fermi surface magnetization expressed by the average magnetic moment of electrons on the Fermi surface. We also carefully excluded other sources of phase shift: the magnetic field error of the superconducting magnets, the internal magnetic fields from the sample magnetization, and the effect of the spin-dependent scattering and exchange field [26–29].

## II. EXPERIMENT DETAILS

$NbSb_2$ single crystals were synthesized by chemical vapor transport with an iodine agent, and the detailed procedure was the same as that in Ref. [19]. For the doped samples, the stoichiometries in the synthesis procedure were adjusted to $Nb_{1-x}(Bi, Fe)_xSb_2$, where x = 0.2, and 0.05 for Bi, and Fe, respectively. For that, we used high-purity Fe powder (99.998%) and ground Bi needle (99.998%). We obtained mm-size samples elongated along the crystal b-axis for all samples. An electron probe X-ray microanalyzer (JXA-8530F, JEOL Ltd, Japan) experiment was performed at 15 keV and 20 nA to verify the actual dopant compositions of the Fe samples. The Fe composition was less than 0.02%, which shows that the actual dopant composition tends to be much smaller than the nominal composition in the synthesis process. Powder X-ray diffraction (XRD) (Rigaku DMAX 2500 diffractometer, Rigaku, Japan) with Cu Kα radiation (λ=1.5406 Å) operated at 40 kV, and 15 mA was conducted to confirm the crystal structure. The XRD spectra of the doped samples are almost the same as that of pristine $NbSb_2$ (see Fig. S1 in Supplemental Material [30]), indicating small doping ratios. The magnetization was measured using a commercial magnetic property measurement system (MPMS, Quantum Design) with a superconducting quantum interference device-vibrating sample magnetometer (SQUID-VSM).

## III. RESULTS AND DISCUSSION

We prepared $NbSb_2$ single crystals with the nonmagnetic and magnetic dopants of Bi and Fe, respectively. The crystal structure of $NbSb_2$ is monoclinic (space group: C12/m1) with centrosymmetry [31], which guarantees the spin degenerate Fermi surface with its nonmagnetic nature. First, we examined the magnetic properties of pristine, Bi-doped, and two Fe-doped $NbSb_2$ single crystals, which are referred to as pristine, Bi, Fe and Fe', respectively. The external magnetic field was applied along the crystal b-axis. Fig. 1(a) shows the magnetic susceptibility, $\chi$, as a function of inverse temperature, $T^{-1}$. For all samples, $\chi$ is negative due to the diamagnetic response of valence electrons. In addition to the diamagnetic response, the linearly increasing $\chi$ with increasing $T^{-1}$ manifests the Langevin-like paramagnetism from the localized magnetic moments [32]. Much steeper slopes in the plots of Fe-doped samples than in those of the other samples indicate that magnetic Fe impurities introduce much more Langevin paramagnetism than nonmagnetic Bi impurities. We fitted the linearly increasing low-temperature (2–100 K) data of $\chi$ with the Curie law $\chi = Np^2\mu_B^2/3k_BT$, where $N$ is the impurity density, $p = g[J(J+1)]^{1/2}$ is the effective number of Bohr magnetons of the impurity, $g$ is the g factor, $J$ is the angular momentum quantum number, and $\mu_B$ is the Bohr magneton. Assuming $Fe^{3+}$ with $g = 2$, $J = 5/2$, and $p = 5.9$, we obtained $N = 5.4$ and $4.4$ ppm for Fe and Fe', respectively. We also measured the field-dependent magnetization $M(B)$ of Fe-doped samples and fitted them with the formula for the field-dependent Langevin paramagnetism $M = NgJ\mu_B B_J(gJ\mu_B B/k_B T)$, where $B_J(x)$ is the Brillouin function. Fig. 1(b) shows $M(B)$ after subtracting the linear diamagnetic signal. Assuming $Fe^{3+}$, we obtained $N = 5.6$ and $4.5$ ppm for Fe and Fe', respectively, which are consistent with the values from the Curie law. We concluded that Fe doping introduces dilute localized magnetic impurities, giving Langevin paramagnetism, while Bi dopants act as nonmagnetic impurities.

Next, we analyzed the dHvA oscillations, $\Delta M$, obtained by subtracting the background magnetization data. Generally, in the three-dimensional spin degenerate system with an ellipsoidal Fermi surface, the dHvA oscillations are given by the LK formula as follows [6,9,19,33].

$$\Delta M \approx -2D_0\sqrt{B}\sum_{p=1}^{\infty}\frac{1}{p^{3/2}}R_T(p)R_D(p)|\cos p\lambda_{as}|\sin\left[2\pi p\left(\frac{F}{B}-\frac{1}{2}\right)+\Theta+p\lambda_s-\frac{\pi}{4}\right], \qquad (1)$$

where $D_0$ is the amplitude of fundamental harmonics without reduction, $R_T = (p\alpha T/B)/\sinh(p\alpha T/B)$ is the temperature reduction factor, $R_D = \exp(-p\alpha T_D/B)$ is the Dinge reduction factor, $\lambda_{as} = (\lambda_\uparrow - \lambda_\downarrow)/2$, $\lambda_s = (\lambda_\uparrow + \lambda_\downarrow)/2$, $\Theta = \frac{\pi}{2}\{1-\text{sign}(\cos\lambda_{as})\}$, $\alpha = 2\pi^2 m_c k_B/e\hbar$, $T_D = \hbar/2\pi k_B \tau$ is the Dingle temperature, and $\lambda_{\uparrow(\downarrow)}$ is the phase from the Fermi surface magnetization of spin up (down) electrons (see Appendix A for the discussion about $\lambda$). We analyzed our data based on Eq. (1).

Fig. 2 shows the analysis of dHvA oscillations. As shown in Fig. 2(a), the oscillation amplitudes of doped samples are relatively small compared to that of the pristine sample, implying a shorter relaxation time for doped samples. We performed a fast Fourier transform (FFT) for the frequency analysis of the dHvA oscillations. Fig. 2(b) shows the FFT spectra. We observed four peaks in the FFT spectra and labeled the peaks as $\zeta$ (~200 T), $\alpha$ (~380 T), β (~700 T), and 2β (~1400 T). The behavior of the FFT spectra is similar to the previous results for NbSb$_2$ [19,34,35]. Overall, the peak positions are not changed by such a tiny doping ratio. Since the β oscillation amplitudes are predominant for all samples, we used the β peaks for the analysis of the oscillations. We applied an FFT bandpass filter to the appropriate frequency range including β peaks (the red shaded area in Fig. 2(b)). The filtered β oscillations are monotonically damped with the increasing inverse field, $1/B$ (see Fig. S2 in Supplemental Material [30]), which is appropriate for applying LK formula.

We extracted the cyclotron mass, $m_c$, of β oscillations by fitting the temperature-dependent oscillation amplitudes with $R_T$ (see Fig. S3 in Supplemental Material [30]). The $m_c$ values of all samples are close to half the electron mass, $0.5m_e$ (listed in Table I). From the doping-independent frequencies and cyclotron masses of β oscillations, we concluded that dilute doping does not significantly change the shape and size of the β Fermi surface. From the field-dependent amplitude, $A(B)$, we introduced $D = A(B)/R_T\sqrt{B}$, which is proportional to $R_D$. Fig. 2(c) shows the Dingle plot, $\ln D$ vs. $B^{-1}$. From the linear slope of the plots, which is proportional to $-\alpha T_D$, we obtained $\tau = 7.8$ ps for the pristine sample and approximately 1 ps for the doped samples (listed in Table I). In contrast to oscillation frequency and cyclotron mass, the values of $\tau$ are dramatically reduced by impurity doping, which means that the dopants have a significant effect on the scattering in the crystal.

We considered the magnetic field error for the phase analysis, which can produce considerable phase error, especially when the applied magnetic field (~ 7 T) is small compared to the frequency of the dHvA oscillatons (~ 700 T). For this reason, we corrected the magnetic field by measuring magnetic field error (see Appendix B). After that, we assigned $N = n + 1/4$ ($n + 3/4$) for $B_{\min(\max)}$, where $B_{\min(\max)}$ is the value of the magnetic field at the minimum (maximum) of the oscillations, and $n$ is a positive integer. With this assignment, we plotted the Landau fan diagram of $1/B_{\min(\max)}$ vs. $N$ (Fig. 2(d)) and obtained the frequency, $F$, from the slopes and the phase, $\gamma$, from the $N$ intercepts. We chose sets of $n$ values to make the values of $\gamma$ lie in the range between $-0.5$ and $0.5$. The obtained $F$ values are close to 700 T for all samples (listed in Table I), consistent with the FFT results of the β peak (~700 T). However, notably, there is a clear phase deviation between the magnetically doped samples (Fe and Fe') and the nonmagnetic counterparts (pristine and Bi) (see the inset of Fig. 2(d)), whose phase is close to the expected value of the LK formula, $-0.125$. The phase deviations from $-0.125$ are $\Delta\gamma = 0.298$ and $0.238$ for Fe and Fe', respectively. We also directly fitted the β oscillations of all samples with the LK formula, and all the parameters, including phase, are similar to the parameters obtained from the amplitude fitting by reduction factors and the Landau fan diagram analysis (see Fig. S2 in Supplemental Material [30]).

We found the phase shift can be explained by the Fermi surface magnetization, whose information is encoded in the phase $\lambda_s$ in Eq. (1), which is elaborated in Appendix A. Conveniently, the relation between $\lambda_s$ and the average magnetic moment of the Fermi surface can be expressed by substituting $\mathcal{D} = m_c/2\pi\hbar^2$ into Eq. (A4) as follows.

$$\lambda_s(B) = \pi \frac{m_c}{m_e} \frac{1}{\mu_B B} \int_0^B <\mu_s(B')>_{E_F} dB' = \pi \frac{m_c}{m_e} \frac{\mu_{eff,s}}{\mu_B}, \qquad (2)$$

where the average $<>_{E_F}$ is taken from the electrons on the Fermi surface with the Fermi energy $E_F$, $\mu_s = (\mu_\uparrow + \mu_\downarrow)/2$ is the average magnetic moment of an electron on the Fermi surface, and $\mu_{\uparrow(\downarrow)}$ is the magnetic moment of an electron on the spin-up (-down) Fermi surface. Note that $\lambda_s(B)$ is proportional to the magnetic energy shift for a field-dependent magnetic moment divided by $B$. We defined the effective average magnetic moment on the Fermi surface as $\mu_{eff,s} = B^{-1} \int_0^B <\mu_s(B')>_{E_F} dB'$, so $\lambda_s$ from a field-dependent $\mu_s(B)$ can be effectively viewed as a result of field-independent $\mu_{eff,s}$. We evaluated $\mu_{eff,s}$ from the phase deviation using the relation $\mu_{eff,s} = (2m_e\Delta\gamma/m_c)\mu_B$. We estimated $\mu_{eff,s} \sim (1.17 \pm 3.93n)\mu_B$ and $(0.98 \pm 4.13n)\mu_B$ for Fe and Fe', respectively, where the positive (negative) value of $\mu_{eff,s}$ indicates decreasing (increasing) magnetic energy. The uncertainty comes from the fact that we can only determine $\Delta\gamma$ (mod 1) in the experiment, and the uncertainty can be resolved by measuring MQOs up to the quantum limit. Unfortunately, the high frequency of β oscillations needs an extremely high field ~ 700 T to observe the quantum limit. The minimal sizes of $\mu_{eff,s}$ are $1.17\mu_B$ and $0.98\mu_B$ for Fe and Fe', respectively, which are still large values considering the magnetic moment of a free electron ≈ $1\mu_B$. Such large $\mu_{eff,s}$ may be attributed to the large effective g factor, which has been usually observed in topological semimetals [20,36–39].

It is noteworthy that the paramagnetic or nonmagnetic nature force $\lambda_s(B=0) = 0$ since the Fermi surface magnetization should be zero in the absence of the applied magnetic field. For pristine and Bi-doped samples, the Fermi surface magnetization is close to zero even in the applied magnetic field, producing $\gamma \approx -0.125$. In contrast, $\gamma = 0.173$ and $0.113$ of Fe-doped samples indicate that the finite Fermi surface magnetization is developed by the applied magnetic field. In Fe-doped samples, the localized paramagnetic impurities are polarized by the applied magnetic fields, which may produce finite Fermi surface magnetization via the interaction between the conduction electrons and localized magnetic impurities.

Fig. 3 shows an intuitive scheme to understand the magnetic moment change by considering the spin-orbit coupling (SOC) and exchange interaction (EI) in the presence of magnetic impurities. The upper panel of Fig. 3 shows the Landau energies of spin up (the solid red line) and down (the solid blue line) states with the Zeeman energy splitting and the average of two energies (the solid light-blue line). The red and blue arrows at each energy level represent the magnetic moments. In the lower panel of Fig. 3, we show the dHvA oscillations of spin up (dashed red line) and down (dashed blue line) electrons and the superposition of the two oscillations (solid black line). First, we discuss the case where there is no SOC or EI (Fig. 3(a)). This situation corresponds to the normal Zeeman splitting in the presence of a magnetic field, giving rise to spins parallel or antiparallel to the direction of the magnetic field. In this case, the magnetic moments of spin up and down electrons compensate each other and produce zero Fermi surface magnetization, consequently, $\lambda_s(B) = 0$, which is shown in the lower panel. The phase difference of spin up and down oscillations, $\lambda_{as} = (\lambda_\uparrow - \lambda_\downarrow)/2$, and the related amplitude modulation, $|\cos \lambda_{as}|$ are also shown. Next, Fig. 3(b) shows the case with SOC and without EI. The direction of spins is tilted by the interaction with the orbital magnetic moments without necessarily being aligned along the direction of the magnetic field. Furthermore, one should consider the orbital angular momentum in addition to the spin angular momentum for the magnetic moment [33,40,41]. The SOC of $NbSb_2$ ~ 100 meV, which is estimated from the gap produced by the SOC in Ref. [31], is much large than the magnetic coupling energy of 7 T ~ 0.4 meV. This means the spin direction is little altered by the magnetic field, which produces almost zero Fermi surface magnetization even in the magnetic field, and $\lambda_s(B) \approx 0$. Finally, looking at the case with SOC and EI (Fig. 3(c)), we find nonzero $\lambda_s(B)$. When we turn on the EI, which is the exchange interaction between the magnetic impurities and the spins on the Fermi surface, the spins tend to be polarized along the polarization direction of the magnetic impurities to minimize the exchange energy, which may be proportional to the polarization strength of the magnetic impurities. This produces asymmetric tilt of spins and nonzero Fermi surface magnetization, which generates $\lambda_s(B) \neq 0$ and the phase deviation in the lower panel.

Additionally, we also investigated other possible sources of the phase shift. First, we considered the field error produced by the internal magnetic field $B$ generated by the sample magnetization. Namely, the magnetic field in the sample should be modified by the sample magnetization to $B = H + 4\pi M$. However, in our case, the magnetization value at the highest field of 7 T is less than $1\times10^{-3}$ $\mu_B$/f.u., which corresponds to $4\pi M = 1.74$ G. Considering our experimental condition $B \sim 5$ T, and the frequency of our dHvA oscillations $\sim 700$ T, the error in phase $\gamma$ from the field error is about $5\times10^{-3}$, which is negligible.

Second, we considered the case of spin-dependent scattering. There have been several previous studies to understand the phase shift in dilute magnetic systems considering the spin-dependent scattering and exchange fields of magnetic impurities, $B_{ex}$. The spin-dependent scattering makes the oscillations from the particular spin dominant and produces the phase shift. We investigated those effects by fitting the amplitude of dHvA effects. In Appendix C, we demonstrated that the spin-dependent scattering and $B_{ex}$ are present in Fe and Fe' samples, and the estimated corresponding phase shifts are $\sim 0.005$ and $0.084$ for Fe and Fe', respectively, which are much smaller than the observed phase shift. Therefore, we excluded the spin-dependent scattering as a dominant origin of the observed phase shifts. Instead, we compared the spin-dependent scattering and $B_{ex}$ in our samples with other samples previously investigated by the spin-dependent scattering model.

As similar analytical results appear in dilute magnetic semiconductors such as Fe-doped HgSe [27] and Cr-doped PbTe [29], we now discuss the physical meaning of the parameters obtained from the spin-dependent scattering analysis. In Table II, we compared the parameters of $B_{ex}$, $\overline{T}_D = (T_{D,\uparrow} + T_{D,\downarrow})/2$, $\delta T_D = (T_{D,\uparrow} - T_{D,\downarrow})/2$, and $m_c$, where $T_{D,\uparrow(\downarrow)}$ is the Dingle temperature of spin up (down) Fermi electrons. The fact that the $B_{ex}$ for Fe-NbSb$_2$ is much smaller compared to those for Fe-HgSe and Cr-PbTe indicates a much smaller magnetic impurity concentration of Fe-NbSb$_2$ because $B_{ex}$ is proportional to the magnetic impurity density. This is the reason why Fe-NbSb$_2$ is paramagnetic rather than semimagnetic, as in the other systems. The average Dingle temperature, $\overline{T}_D$, for Fe-NbSb$_2$ is slightly smaller than those for Fe-HgSe and Cr-PbTe despite the much smaller magnetic impurity concentration, and the difference in Dingle temperature, $\delta T_D$, of Fe-NbSb$_2$ is close to Cr-PbTe. Since $T_D$ is proportional to the scattering rate, we expect a much more scattering rate for a magnetic impurity in Fe-NbSb$_2$, which is well explained by the much larger density of states on the Fermi surface. The density of states, approximated by $m_c^{3/2}$, is more than 50 times larger for Fe-NbSb$_2$ than for Fe-HgSe and Cr-PbTe. This can allow considerable scattering rates and interaction between Fermi electrons and magnetic impurities, which may lead to a sufficiently large effective average magnetic moment of the Fermi elections, $\mu_{eff,s} \geq 1\mu_B$. However, clarifying the detailed mechanism explaining the relation between Fermi surface magnetization and scattering with magnetic impurities needs further studies.

## IV. CONCLUSION

In conclusion, we carefully analyzed the phase of MQOs in the magnetically doped semimetal Fe-NbSb$_2$ compared to the nonmagnetic semimetals pristine NbSb$_2$ and Bi-doped NbSb$_2$. Although we adopted a small doping ratio to preserve the shape of the Fermi surface, we observed phase shifts of $\Delta\gamma = 0.298$ and $0.238$ for Fe-NbSb$_2$ and no phase shift for Bi-doped NbSb$_2$. To interpret the unusual phase shift due to magnetic doping, we investigated possible sources of phase shift, such as magnetic field error, spin-dependent scattering, and exchange field. All reported explanations cannot explain our results. Therefore, we suggested the Fermi surface magnetization as the origin of the unusual phase shift and developed a methodology to investigate Fermi surface magnetization from the phase of MQOs. This methodology for phase analysis can be used for various magnetic materials, and our results provide a new perspective in understanding the interaction between Fermi surface electrons and localized magnetic moments.

## ACKNOWLEDGMENTS

This work was supported by a National Research Foundation of Korea (NRF) grant (No. 2020R1A2C3008044).

## APPENDIX A: Phase from the Fermi surface magnetization

For a nondegenerate and three-dimensional ellipsoidal Fermi surface, the phase of MQOs $\gamma$ is given by [33]

$$2\pi\gamma = \lambda - \pi - \pi/4, \tag{A1}$$

where $\lambda$ is the phase from the magnetic property of the Fermi surface. $\lambda$ can be expressed most conveniently with the field-dependent Fermi surface magnetization, which is directly derived from Eq. (1) of Ref. [6], as

$$\lambda(B) = 2\pi\phi_0 B^{-1} \frac{\partial}{\partial E_F} \int_0^B M(B', E_F) dB', \tag{A2}$$

where $\phi_0$ is the magnetic flux quantum, $E_F$ is the Fermi energy, and $M(E_F)$ is the two-dimensional magnetization with $E_F$. Here, $M$ is a continuous function of $E_F$, which is defined from the semiclassical free energy $G_{semi}$ (continuous function of $E_F$) by $M = -\partial G_{semi}/\partial B$. Note that in a three-dimensional system, $M(E_F)$ should be the magnetization contributed by the extremal cross-section of the Fermi surface. For a spin-degenerate Fermi surface, the phase from the magnetic property $\lambda_s$ is given by the average of phases from spin up $\lambda_\uparrow$ and down $\lambda_\downarrow$ as [33]

$$\lambda_s(B) = \frac{\lambda_\uparrow(B) + \lambda_\downarrow(B)}{2} = 2\pi\phi_0 B^{-1} \frac{\partial}{\partial E_F} \int_0^B M_s(B', E_F) dB', \tag{A3}$$

where the average magnetization $M_s = (M_\uparrow + M_\downarrow)/2$ defined with the magnetization of the spin up (down) Fermi pocket, $M_{\uparrow(\downarrow)}$. $\lambda_s$ can alternatively be expressed by the average magnetic moment of an electron on the Fermi surface $\mu_s = (\mu_\uparrow + \mu_\downarrow)/2$ as follows.

$$\lambda_s(B) = 2\pi\phi_0 B^{-1} \frac{\partial}{\partial E_F} \int_0^B \int^{E_F} \mathcal{D}(E)\mu_s(B', E) dE dB' = 2\pi\phi_0 \mathcal{D}(E_F) B^{-1} \int_0^B <\mu_s(B')>_{E_F} dB', \tag{A4}$$

where $\mathcal{D}$ is the density of states and the average $<>_{E_F}$ is taken from the electrons at the Fermi energy.

**APPENDIX B: PHASE CORRECTION CONSIDERING MAGNETIC FIELD ERROR**

The magnetic field error caused by the conventional superconducting magnet is summarized in Application Note 1070-207 of Quantum Design [42]. We measured the magnetization for the dHvA oscillations with "stable at each field" using the persistent mode. The possible field errors in the persistent mode come from inconsistency between the calibrated field from the measured current before entering the persistent mode and the real magnetic field. The typical origins are the trapped flux in the superconducting magnet and the escaped flux. The error value depends on the field sweep history and the magnet status.

We obtained the magnetic field error by comparing the theoretical and measured magnetic moment of a Pd standard sample provided by Quantum Design, which is shown in Fig. 4(a). Before the measurement, the field was set to zero and rapidly set to 7 T (500 Oe/sec). Subsequently, the measurement was conducted with decreasing and increasing magnetic fields. Since the dHvA measurements were conducted with decreasing magnetic fields, we used the magnetic field error measured in decreasing magnetic fields.

Given that the magnetic field error, the dHvA oscillations can be rewritten as $-\sin(2\pi F/B+\phi) = -\sin(2\pi F/(B_{rep}+\Delta B)+\phi)$ where $B_{rep}$ is the reported field value and $\Delta B$ is the magnetic field error defined as $B = B_{rep}+\Delta B$. For the magnetic field error much smaller than the real field, $-\sin(2\pi F/(B_{rep}+\Delta B)+\phi) = -\sin(2\pi F/B_{rep} -$ $(2\pi F/B_{rep}^2)\Delta B+\phi)$ is satisfied, and the phase error is $\Delta\gamma_{FE} = \Delta\phi_{FE}/2\pi = -(F/B_{rep}^2)\Delta B$. By $\Delta\gamma_{FE}$, the phase of dHvA oscillations extracted from the Landau fan diagram or direct fitting can differ from the intrinsic value. In the magnetic field range of 3.5 T - 7 T, we corrected the magnetic field values. The phase from the Landau fan diagram after the correction is −0.109 (−0.091) for pristine (Bi), which is closer to the expected value of −0.125 than the value of −0.179 (−0.162) obtained before the correction.

**APPENDIX C: ESTIMATION OF THE POSSIBLE PHASE DEVIATION FROM THE SPIN-DEPENDENT SCATTERING**

ln $D$ and the phase deviation $\Delta\theta_{SD}$ in the presence of spin-dependent scattering is given by [27]

$$\ln D = \ln D_0 - \frac{\alpha \overline{T}_D}{B} + \frac{1}{2}\ln[e^{-2\alpha\delta T_D/B} + e^{+2\alpha\delta T_D/B} + 2\cos(2\lambda_{as})], \quad (C1)$$

and

$$\Delta\theta_{SD} = \arctan\left[\tan(\lambda_{as})\left(\frac{1-e^{-2\alpha\delta T_D/B}}{1+e^{-2\alpha\delta T_D/B}}\right)\right], \quad (C2)$$

where $\overline{T}_D = (T_{D,\uparrow} + T_{D,\downarrow})/2$, $\delta T_D = (T_{D,\uparrow} - T_{D,\downarrow})/2$, $T_{D,\uparrow(\downarrow)}$ is the Dingle temperature of spin up (down) Fermi surface, $\lambda_{as} = = \pi g_{eff} m_c/2m_e$, and $g_{eff}$ is the effective g factor. In the presence of exchange fields $B_{ex}$, $g_{eff} = g + B_{ex}/B$, where $g$ is the g factor without $B_{ex}$. From Eqs. (1) and (C1), the below relation is satisfied.

$$\frac{A(2)}{A(1)} = \frac{1}{2^{3/2}} \frac{|\cos(2\lambda_{as})| R_T(2) e^{-2\alpha\overline{T}_D/B} \sqrt{e^{-4\alpha\delta T_D/B} + e^{+4\alpha\delta T_D/B} + 2\cos(4\lambda_{as})}}{|\cos(\lambda_{as})| R_T(1) e^{-\alpha\overline{T}_D/B} \sqrt{e^{-2\alpha\delta T_D/B} + e^{+2\alpha\delta T_D/B} + 2\cos(2\lambda_{as})}} \quad (C3)$$

where $A(p)$ is the amplitude of $p$th harmonics. First, we evaluated $\lambda_{as} = 0.56\pi$ (mod $2\pi$) for pristine from Eq. (C3), setting $\delta T_D$, $B_{ex}$, = 0. In the evaluation, $A(1)$ and $A(2)$ were taken at 6.5 T and $\overline{T}_D$ and $R_T$ are evaluated with $\tau =$ 7.8 ps and $m_c = 0.514 m_e$ in Table I. Solving Eq. (C3) gives four values of $\lambda_{as}$ (mod $2\pi$) in each range $[0, 0.5\pi)$, $[0.5\pi, \pi)$, $[\pi, 1.5\pi)$, and $[1.5\pi, 2\pi)$, but the range $[0.5\pi, \pi)$ was chosen following the argument in [19]. Subsequently, ln $D_0 = -8.7$ was evaluated for pristine by the relation ln $D = \ln(2|\cos\lambda_{as}|D_0)$ at $1/B = 0$, which is the intercept of ln $D - 1/B$ plot in Fig. 4(c). We also obtained ln $D_0 = -8.7$ and −8.6 for Fe and Fe', respectively, from the relation of $D_0 \propto 1/m_c$, which is valid for constant $F$ [9].

With fixed ln $D_0$, we fitted the amplitudes of Fe and Fe' with Eq. (C1). In the fitting procedure, we imposed Eq. (C3) with $A(2)/A(1)$ values at 6.5 T as a constraint on the fitting parameters. Fig. 5(a) shows the amplitudes of dHvA oscillations and fitted curves by Eq. (C1) for Fe and Fe'. We obtained $\overline{T}_D$ = 1.2 and 1.1 K, $\delta T_D$ = 0.05 and 0.05 K, $g$ = 2.15 and 2.27, and $B_{ex}$ = −0.23 and −0.72 T for Fe and Fe', respectively. Note that there are multiple choices of $g$, but we chose the value to make $\lambda_{as}$ in the range [0.5π, π). Any choice of $g$ does not affect the estimation of $\Delta\theta_{SD}$. From the parameters, we simulated $\Delta\gamma_{SD} = \Delta\theta_{SD}/2\pi$ as a function of $1/B$ by Eq. (C2), which is shown in Fig. 5(b). The average $\Delta\gamma_{SD}$ are ~0.11 and ~0.15 for Fe and Fe', respectively. However, the phase obtained from the Landau fan diagram or direct fitting of dHvA oscillations would not be the same as $\Delta\gamma_{SD}$, since $\Delta\gamma_{SD}$ is field-dependent. Therefore, we simulated dHvA oscillations by the parameters obtained from the spin-dependent scattering analysis and plotted Landau fan diagrams. Fig. 5(c) shows the Landau fan diagrams and their intercepts, which show that the phase derivations are −0.005 and −0.084 for Fe and Fe', respectively. The phase deviations from the spin-dependent scattering are much smaller than the observed phase deviations, which are $\Delta\gamma$ = 0.298 and 0.238 for Fe and Fe', respectively. Note that the sign of phase deviation from the spin-dependent scattering cannot be determined by the analysis in this Appendix. As a result, the phase deviations can also be +0.005 and +0.084 for Fe and Fe', respectively.

TABLE I. Physical parameters obtained from the dHvA oscillations for pristine, Bi-doped, and Fe-doped NbSb$_2$: oscillation frequency $F$, cyclotron mass $m_c$, relaxation time $\tau$, and phase $\gamma$.

| Sample | $F$ (T) | $m_c$ ($m_e$) | $\tau$ (ps) | $\gamma$ |
|---|---|---|---|---|
| Pristine | 705 | 0.514(5) | 7.8(1) | −0.109(5) |
| Bi | 709 | 0.513(3) | 0.94(1) | −0.091(8) |
| Fe | 705 | 0.508(7) | 1.06(1) | 0.173(5) |
| Fe' | 705 | 0.485(3) | 1.18(3) | 0.113(1) |

The values in parentheses denote the uncertainty of the last digit. $F$ are accurate up to the first decimal place.

TABLE II. Comparison of physical parameters of Fe-doped NbSb$_2$, Fe-doped HgSe, and Cr-doped PbTe: exchange field $|B_{ex}|$, average Dingle temperature $\overline{T}_D$, difference of Dingle temperature $\delta T_D$, and cyclotron mass $m_c$.

| Sample | $|B_{ex}|$ (T) | $\overline{T}_D$ (K) | $\delta T_D$ (K) | $m_c$ ($m_e$) |
|---|---|---|---|---|
| NbSb$_2$(Fe) | 0.23(8) | 1.2(1) | 0.05(1) | 0.508(7) |
| Hg$_{1-x}$Fe$_x$Se [27] | 36(1) | 3.0(2) | 0.3(1) | 0.067(1) |
| PbTe(Cr) [29] | 13.0(2) | 3.1(1) | 0.03(1) | 0.0808 |

For Hg$_{1-x}$Fe$_x$Se and PbTe(Cr), $|B_{ex}|$ is obtained by $g_1$ values in Ref. [27,29], which indicates $B_{ex}$ in saturating magnetization, namely, $B_{ex} = g_1 B_J$.

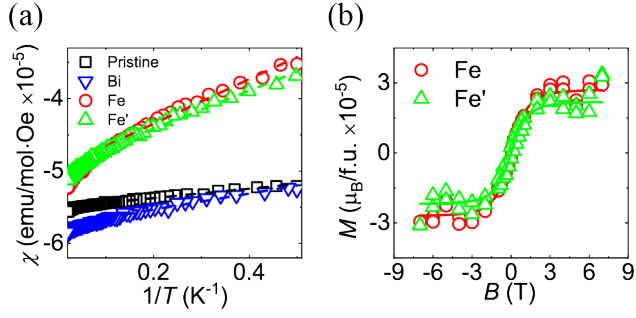

FIG. 1. Temperature- and field-dependent magnetic properties of pristine and doped $NbSb_2$. (a) Magnetic susceptibility versus inverse temperature ($\chi$ vs. $T^{-1}$) measured at a 1 T field. The open symbols represent the experimental data, and the dashed lines represent the linearly fitted curves. The slopes were analyzed with the Curie law $\chi = Np^2\mu_B^2/3k_BT$, from which we obtained the magnetic impurity concentration: $N$ = 5.4 and 4.4 ppm for Fe and Fe', respectively. (b) Magnetization versus magnetic field ($M$ vs. $B$) of Fe-doped $NbSb_2$. The open symbols represent the experimental data, and the solid lines represent the fitted curve by the Langevin paramagnetism, $M = NgJ\mu_B B_J(gJ\mu_B B/k_B T)$, where the obtained magnetic impurity concentrations are $N$ = 5.6 and 4.5 ppm for Fe and Fe', respectively, which are in good agreement with the values obtained from the Curie law.

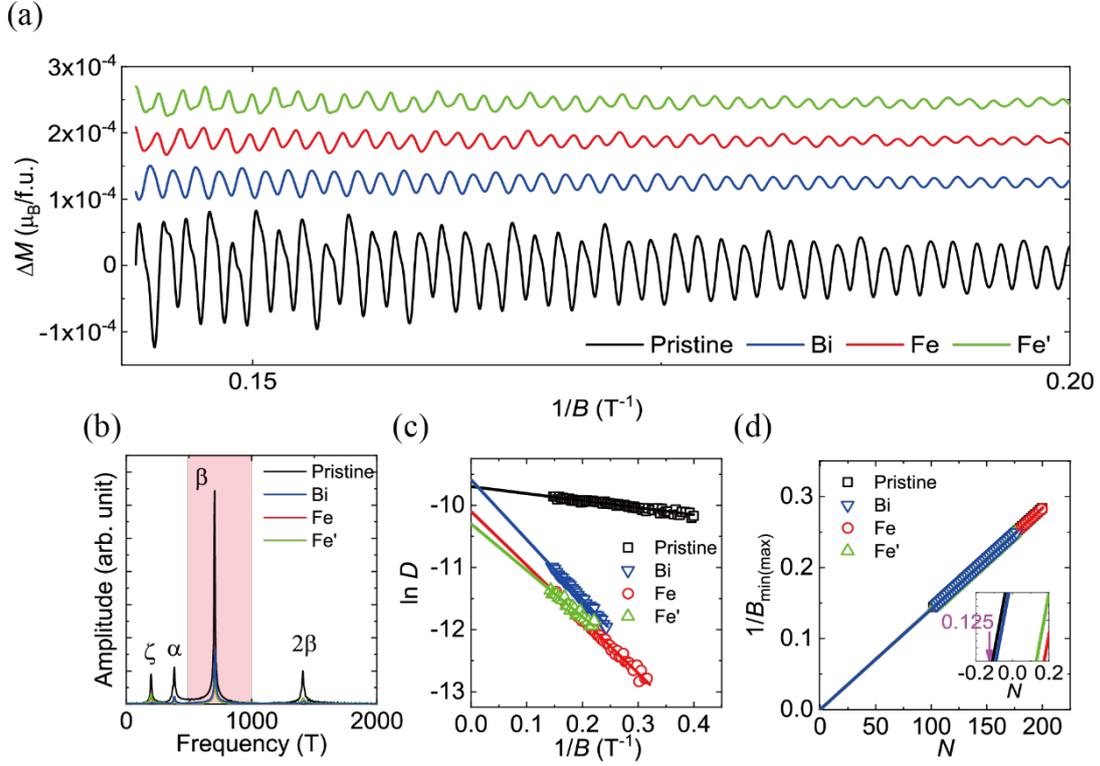

FIG. 2. dHvA effects for pristine and doped NbSb$_2$. (a) Oscillatory signals of magnetizations, $\Delta M$, obtained after subtracting the background magnetization data. (b) FFT spectra of $\Delta M$, where the peaks are labeled $\zeta$ (~ 200 T), $\alpha$ (~ 380 T), $\beta$ (~ 700 T), and $2\beta$ (~1400 T). The red shaded area is the bandpass FFT filter range to investigate the $\beta$ oscillations separately. (c) Dingle plots (ln $D$ vs. $B^{-1}$), where $D = A(B)/R_T\sqrt{B}$, from which the relaxation time, $\tau$, is obtained. (d) Landau fan diagrams of $\beta$ oscillations, in which $N = n + 1/4$ ($n + 3/4$) is assigned to the minimum (maximum) of the $\beta$ oscillations and $n$ is a positive integer. The open symbols represent the experimental data, and the solid lines represent the linearly fitted curves and their extrapolations. The inset shows the extrapolated lines near the $N = 0$ point. The obtained parameters are listed in Table I.

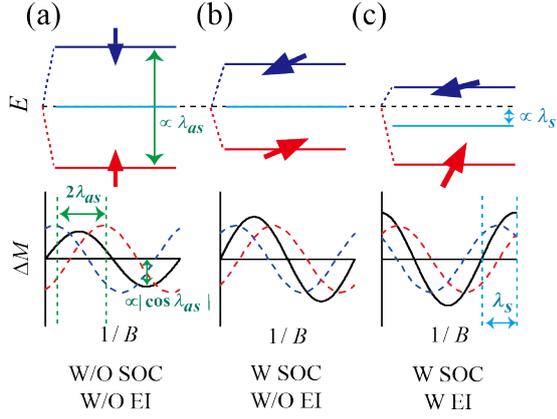

FIG. 3. Illustration of the Landau levels and the corresponding dHvA oscillations. In the upper panel, the solid red and blue arrows show the spin up and down magnetic moment vectors, respectively, with the Landau energies (the solid lines) shifted by the Zeeman energy splitting. The solid light-blue lines are the average energy levels of the spin up and down Landau energies. In the lower panel, the dashed red and blue lines show the spin up and down oscillations, respectively, and the solid black lines show the superpositions of the two oscillations. (a) Case without SOC and EI. The phase difference of spin up and down oscillations, $\lambda_{as} = (\lambda_\uparrow - \lambda_\downarrow)/2$, and the related amplitude modulation, $|\cos \lambda_{as}|$, are shown. There is no phase deviation since the average magnetic moment is zero. (b) Case with SOC and without EI. The direction of spins is tilted by the interaction with the orbital magnetic moments without necessarily being aligned along the direction of the magnetic field. The average magnetic moment is zero, and there is no phase deviation. (c) Case with SOC and EI. The conduction electrons on the Fermi surface interact with the polarized magnetic impurities, which may produce a finite average magnetic moment, leading to the phase deviation, $\lambda_s \neq 0$.

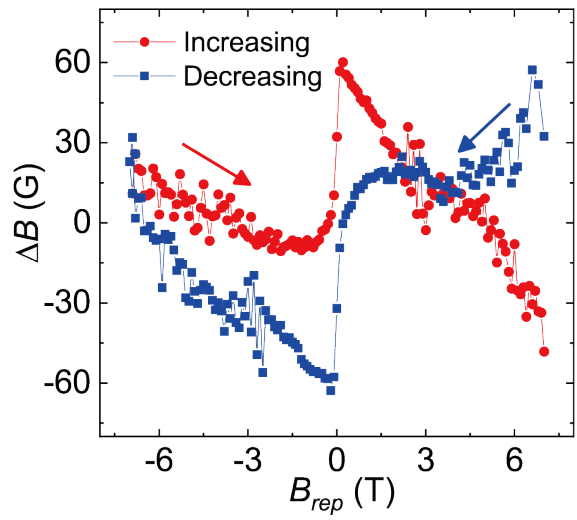

FIG. 4. The field error estimated by comparing the theoretical and experimentally measured magnetic moment of the Pd standard sample. The blue square and red circle symbols indicate the phase errors produced by decreasing and increasing the magnetic field, respectively.

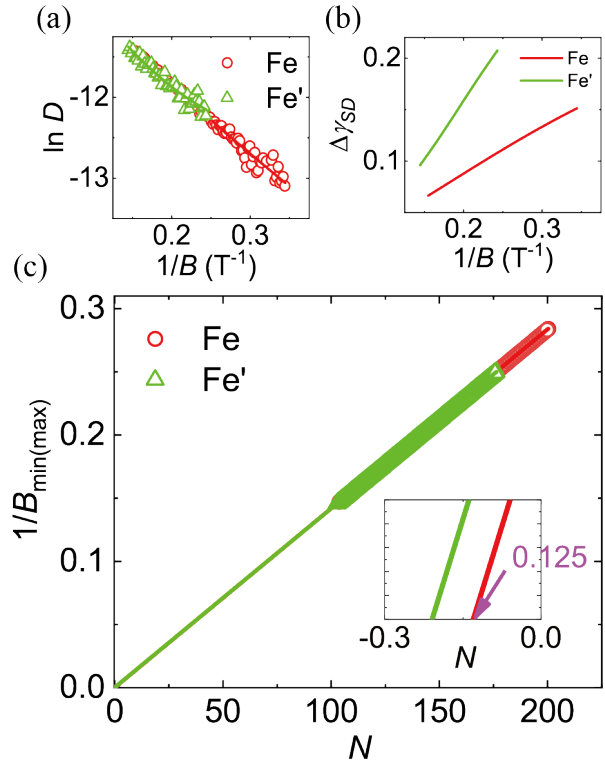

FIG. 5. Effects of the spin-dependent scattering in Fe-doped NbSb$_2$. (a) ln $D$ vs. $B^{-1}$ plots (the open symbols) and fitted curves (the solid lines) by Eq. (C1). The parameters obtained by the fitting are listed in Table II. (b) Simulated phase deviation from the fitting parameters. (c) Landau fan diagrams simulated by the fitting parameters (the open symbols) and the linear extrapolations (the solid lines). The phase deviations from the value of −0.125 are −0.005 and −0.084 for Fe and Fe', respectively.

# REFERENCES

[1] V. L. Schubnikow and W. J. de Haas, *Magnetische Widerstandsvergrösserung in Einkristallen von Wismut Bei Tiefen Temperaturen*, Proc. R. Neth. Acad. Arts. Sci. **33**, 130 (1930).

[2] E. M. Lifshits and A. M. Kosevich, *Theory of the Shubnikov-de Haas Effect*, J. Phys. Chem. Solids **4**, 1 (1958).

[3] E. N. Adams and T. D. Hol, *Quantum Theory of Transverse Galvano-Magnetic Phenomena*, J. Phys. Chem. Solids **10**, 254 (1959).

[4] W. J. de Haas and P. M. van Alphen, *The Dependence of the Susceptibility of Diamagnetic Metals upon the Field*, Proc. R. Neth. Acad. Arts. Sci. **33**, 1106 (1930).

[5] I. M. Lifshitz and A. M. Kosevich, *Theory of Magnetic Susceptibility in Metals at Low Temperatures*, J. Exp. Theor. Phys. **2**, 636 (1956).

[6] Y. Gao and Q. Niu, *Zero-Field Magnetic Response Functions in Landau Levels*, Proc. Natl. Acad. Sci. U.S.A. **114**, 7295 (2017).

[7] A. Alexandradinata and L. Glazman, *Semiclassical Theory of Landau Levels and Magnetic Breakdown in Topological Metals*, Phys. Rev. B **97**, 144422 (2018).

[8] L. M. Roth, *Semiclassical Theory of Magnetic Energy Level and Magnetic Susceptibility of Bloch Electrons*, Phys. Rev. **145**, 434 (1966).

[9] D. Shoenberg and F. R. S., *Magnetic Oscillations in Metals* (Cambridge University Press, Cambridge, 1984).

[10] L. Fang, Y. Jia, D. J. Miller, M. L. Latimer, Z. L. Xiao, U. Welp, G. W. Crabtree, and W. K. Kwok, *Catalyst-Free Growth of Millimeter-Long Topological Insulator $Bi_2Se_3$ Nanoribbons and the Observation of the π-Berry Phase*, Nano Lett. **12**, 6164 (2012).

[11] A. A. Taskin, S. Sasaki, K. Segawa, and Y. Ando, *Manifestation of Topological Protection in Transport Properties of Epitaxial $Bi_2Se_3$ Thin Films*, Phys. Rev. Lett. **109**, 066803 (2012).

[12] S. W. Kim, H. Kim, J. K. Kim, W. S. Noh, J. Kim, K. M. Kim, K. S. Kim, J. S. Kim, J. H. Park, and M. H. Jung, *Effect of Antiferromagnetic Order on Topological Electronic Structure in Eu-Substituted $Bi_2Se_3$ single Crystals*, APL Mater. **8**, 111108 (2020).

[13] J. H. Jun, J. Kim, S. W. Kim, and M. H. Jung, *Signature of Topological States in Antiferromagnetic Sm-Substituted $Bi_2Te_3$*, Sci. Rep. **10**, 9615 (2020).

[14] S. W. Kim and M. H. Jung, *Electronic Properties of $Gd_xBi_{2-x}Se_3$ Single Crystals Analyzed by Shubnikov-de Haas Oscillations*, Appl. Phys. Lett. **112**, 202401 (2018).

[15] S. Ji, S. E. Lee, and M. H. Jung, *Berry Paramagnetism in the Dirac Semimetal $ZrTe_5$*, Commun. Phys. **4**, 265 (2021).

[16] T. Liang, Q. Gibson, M. N. Ali, M. Liu, R. J. Cava, and N. P. Ong, *Ultrahigh Mobility and Giant Magnetoresistance in the Dirac Semimetal $Cd_3As_2$*, Nat. Mater. **14**, 280 (2015).

[17] L. P. He, X. C. Hong, J. K. Dong, J. Pan, Z. Zhang, J. Zhang, and S. Y. Li, *Quantum Transport Evidence for the Three-Dimensional Dirac Semimetal Phase in $Cd_3As_2$*, Phys. Rev. Lett. **113**, 246402 (2014).

[18] Z. J. Xiang, D. Zhao, Z. Jin, C. Shang, L. K. Ma, G. J. Ye, B. Lei, T. Wu, Z. C. Xia, and X. H. Chen, *Angular-Dependent Phase Factor of Shubnikov-de Haas Oscillations in the Dirac Semimetal $Cd_3As_2$*, Phys. Rev. Lett. **115**, 226401 (2015).

[19] S.-E. Lee, Myeong-jun Oh, Sanghyun Ji, Jinsu Kim, Jin-Hyeon Jun, Woun Kang, Younjung Jo, and Myung-Hwa Jung, *Orbit Topology Analyzed from π Phase Shift of Magnetic Quantum Oscillations in Three-Dimensional Dirac Semimetal*, Proc. Natl. Acad. Sci. U.S.A. **118**, e2023027118 (2021).

[20] Y. Liu et al., *Zeeman Splitting and Dynamical Mass Generation in Dirac Semimetal $ZrTe_5$*, Nat. Commun. **7**, 12516 (2016).

[21] J. Wang, J. Niu, B. Yan, X. Li, R. Bi, Y. Yao, D. Yu, and X. Wu, *Vanishing Quantum Oscillations in Dirac Semimetal $ZrTe_5$*, Proc. Natl. Acad. Sci. U.S.A. **115**, 9145 (2018).

[22] G. P. Mikitik and Y. v Sharlai, *Manifestation of Berry's Phase in Metal Physics*, Phys. Rev. Lett. **82**, 2147 (1999).

[23] G. P. Mikitik and Y. v Sharlai, *The Phase of the de Haas-van Alphen Oscillations, the Berry Phase, and Band-Contact Lines in Metals*, Low Temp. Phys. **33**, 586 (2007).

[24] Y. Zhang, Y. W. Tan, H. L. Stormer, and P. Kim, *Experimental Observation of the Quantum Hall Effect and Berry's Phase in Graphene*, Nature **438**, 201 (2005).

[25] A. A. Taskin and Y. Ando, *Berry Phase of Nonideal Dirac Fermions in Topological Insulators*, Phys. Rev. B **84**, 035301 (2011).

[26] S. Xi Zhang et al., *Spin-Dependent Scattering of Conduction Electrons in $Cd_{3-x-y}Zn_xMn_yAs_2$ Alloys*, Semicond. Sci. Technol **6**, 619 (1991).

[27] M. Vaziri and R. Reifenberger, *Spin-Dependent Scattering of Conduction Electrons in Diluted Magnetic Semiconductors: $Hg_{1-x}Fe_xSe$*, Phys. Rev. B **32**, 3921 (1985).

[28] R. Laiho, K. G. Lisunov, E. Lähderanta, V. N. Stamov, and V. S. Zakhvalinskii, *Magnetization and Shubnikov-de Haas Effect in Diluted Magnetic Semiconductors $(Cd_{1-x-y}Zn_xMn_y)_3As_2$*, J. Appl. Phys. **81**, 5151 (1997).

[29] B. A. Akimov, N. A. Lvova, and L. I. Ryabova, *Quantum Oscillatory Properties of the Semimagnetic Semiconductor PbTe(Cr)*, Phys. Rev. B **58**, 10430 (1998).

[30] *See Supplemental Material at " " for the XRD data, the fitted dHvA oscillations by the LK formula, and the temperature-dependent amplitude of the dHvA oscillations.*

[31] C. Xu, J. Chen, G. X. Zhi, Y. Li, J. Dai, and C. Cao, *Electronic Structures of Transition Metal Dipnictides $XPn_2$ (X=Ta, Nb; Pn= P, As, Sb)*, Phys. Rev. B **93**, 195106 (2016).

[32] Charles Kittel, *Introduction To Solid State Physics*, 8th ed. (John Wiley & Sons, 2005).

[33] A. Alexandradinata, C. Wang, W. Duan, and L. Glazman, *Revealing the Topology of Fermi-Surface Wave Functions from Magnetic Quantum Oscillations*, Phys. Rev. X **8**, 011027 (2018).

[34] K. Wang, D. Graf, L. Li, L. Wang, and C. Petrovic, *Anisotropic Giant Magnetoresistance in $NbSb_2$*, Sci. Rep. **4**, 7328 (2014).

[35] L. Guo et al., *Extreme Magnetoresistance and SdH Oscillation in Compensated Semimetals of $NbSb_2$ Single Crystals*, J. Appl. Phys. **123**, 155103 (2018).

[36] R. Bi, Z. Feng, X. Li, J. Niu, J. Wang, Y. Shi, D. Yu, and X. Wu, *Spin Zero and Large Landé g-Factor in $WTe_2$*, New J. Phys. **20**, 063026 (2018).

[37] M. H. Cohen and E. I. Blount, *The g-Factor and de Haas-van Alphen Effect of Electrons in Bismuth*, Philosophical Magazine **5**, 115 (1960).

[38] J. Cao et al., *Landau Level Splitting in $Cd_3As_2$ under High Magnetic Fields*, Nat. Commun. **6**, 7779 (2015).

[39]  J. Xiong, S. K. Kushwaha, T. Liang, J. W. Krizan, M. Hirschberger, W. Wang, R. J. Cava, and N. P. Ong, *Evidence for the Chiral Anomaly in the Dirac Semimetal Na$_3$Bi*, Science **350**, 413 (2015).

[40]  E. I. Blount, *Bloch Electrons in a Magnetic Field*, Phys. Rev. **126**, 1636 (1962).

[41]  M.-C. Chang and Q. Niu, *Berry Phase, Hyperorbits, and the Hofstadter Spectrum: Semiclassical Dynamics in Magnetic Bloch Bands*, Phys. Rev. B **53**, 7010 (1996).

[42]  Quantum Design, Application Note 1070-207: Using PPMS Superconducting Magnets at Low Fields, 2009.